\documentclass[manuscript]{acmart}
\usepackage{kotex}
\usepackage{xcolor}
\usepackage{multirow}

\AtBeginDocument{%
  \providecommand\BibTeX{{%
    \normalfont B\kern-0.5em{\scshape i\kern-0.25em b}\kern-0.8em\TeX}}}

\setcopyright{acmcopyright}
\copyrightyear{2022}
\acmYear{2022}
\acmDOI{XXXXXXX.XXXXXXX}

\acmConference[AutoUI '22]{14th international ACM conference on automotive user interfaces}{September 17--20, 2022}{Seoul, South Korea}

\acmPrice{15.00}
\acmISBN{978-1-4503-XXXX-X/18/06}

\begin{document}

\title{Affective Role of the Future Autonomous Vehicle Interior}

\author{Taesu Kim} 
\email{tskind77@kaist.ac.kr}
\orcid{0000-0002-4846-1805}
\affiliation{%
  \institution{Korea Advanced Institute for Science and Technology}
  \city{Daejeon}
  \country{Korea}
  \postcode{34141}
}

\author{Gyunpyo Lee}
\email{gyunpyolee@kaist.ac.kr}
\affiliation{%
  \institution{Korea Advanced Institute for Science and Technology}
  \city{Daejeon}
  \country{Korea}
  \postcode{34141}
}

\author{Jiwoo Hong}
\email{jwhong10@kaist.ac.kr}
\orcid{}
\affiliation{%
  \institution{Korea Advanced Institute for Science and Technology}
  \city{Daejeon}
  \country{Korea}
  \postcode{34141}
}

\author{Hyeon-Jeong Suk}
\email{color@kaist.ac.kr}
\orcid{0000-0002-0199-6183}
\affiliation{%
  \institution{Korea Advanced Institute for Science and Technology}
  \city{Daejeon}
  \country{Korea}
  \postcode{34141}
}

\renewcommand{\shortauthors}{Kim et al.}

\begin{abstract}

Recent advancements in autonomous technology allow for new opportunities in vehicle interior design. Such a shift in in-vehicle activity suggests vehicle interior spaces should provide an adequate manner by considering users' affective desires. Therefore, this study aims to investigate the affective role of future vehicle interiors. Thirty-one participants in ten focus groups were interviewed about challenges they face regarding their current vehicle interior and  expectations they have for future vehicles. Results from content analyses revealed the affective role of future vehicle interiors. Advanced exclusiveness and advanced convenience were two primary aspects identified. The identified affective roles of each aspect are a total of eight visceral levels, four visceral levels each, including  focused, stimulating, amused, pleasant, safe, comfortable, accommodated, and organized. We expect the results from this study to lead to the development of affective vehicle interiors by providing the fundamental knowledge for developing conceptual direction and evaluating its impact on user experiences.

\end{abstract}

\begin{CCSXML}
<ccs2012>
   <concept>
       <concept_id>10003120.10003121.10003126</concept_id>
       <concept_desc>Human-centered computing~HCI theory, concepts and models</concept_desc>
       <concept_significance>500</concept_significance>
       </concept>
 </ccs2012>
\end{CCSXML}

\ccsdesc[500]{Human-centered computing~HCI theory, concepts and models}

\keywords{Affective Design, Autonomous Vehicle, User Expectation, Vehicle Interior}

\maketitle

\section{Introduction}
The vehicle industry is undergoing rapid innovation as a result of technological advancement. Passenger vehicles are now equipped with intelligent devices that continuously assist drivers. For example, the advanced driver assistance systems (ADAS) that assist the driver in driving, such as rear collision warning, blind-spot detection, and park assist, are essential in passenger vehicles \cite{WhatisAD26}. Accordingly, sensitively designed infotainment systems accommodate a more natural user experience and ease in operation and interaction between passengers and the vehicle. In this circumstance, the general user perception of vehicles as a means of transportation changes to viewing vehicales as smart objects \cite{eichler2006car}. 

\begin{figure}[t]
    \centering
    \includegraphics[width = \textwidth]{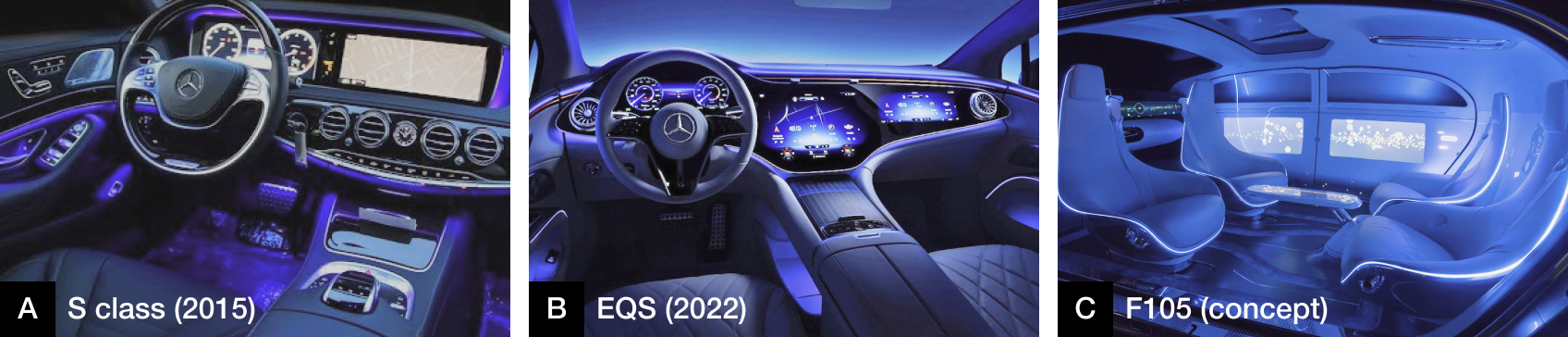}
    \caption{Changes in the vehicle interior design according to the advancement of technologies on the vehicle manufacturer: Case of Mercedes Benz. (a) S class launched in 2015; includes a lot of elements mainly designed for a stylish interior, (b) EQS launched in 2022; removed buttons and equipped with large infotainment display which attatched Mercedes Benz user experience system, (c) F015 concept car; showcases a possible future interior for fully autonomous vehicles.
    %하나의 vehicle manufactur내 자동차 기술 진보에 따른 자동차 인테리어 구조의 변화: Mercedes Benz를 중심으로. (a) 차량의 스타일링적 요소가 다량 포함된 2015년 S class, (b) 미래 지향적 MBUX (Mercedes-Benz User Experience)가 탑재된 2022년 EQS, (c) 완전 자율주행 차량의 미래 공간을 제안하고 있는 F015 concept
    }
    \label{fig:interiorinno}
    \vspace{-5mm}
\end{figure}

Accelerating vehicle innovation with the development and improvement of autonomous technologies will expand current design innovations centered on existing vehicle infotainment systems to the entire vehicle interior spaces (Figure \ref{fig:interiorinno}). The core benefit of autonomous driving technology is that it relieves drivers of the burden of driving and allows for time off while on the road. Therefore, the traditional activity of the driver is substituted with other activities such as entertainment or leisure and more subdivided into experience levels \cite{trommer2016autonomous}. In addition, autonomous driving technology allows for increased and diversified social activity between the driver and passenger(s) \cite{lee2020exploring}. In line with this change, Nissan developed a new design for autonomous vehicle interiors in the 2015 IDS concept car \cite{NissanID32}. They proposed a vehicle interior that blurs the boundary between driver and passenger by introducing front-row seats that rotate when shifted into autonomous mode. The development of autonomous driving technology transforms user perceptions. Accordingly, it opens up new design opportunities to propose innovative vehicle interior spaces \cite{sun2021shaping}.  

Vehicle manufacturers are faced with more consideration of in-depth user activities in their vehicle interior design \cite{zhu2021acceptance, bernhart2021purpose}. For example, Toyota presented the multi-purpose mobility e-palette concept at CES 2018. It is a barrier-free interior platform that can freely design interiors based on customer needs, suggesting an era in which user-oriented vehicle interior design becomes possible \cite{ePalette47}. Furthermore, from the 2018 Paris Motor Show, Renault subdivided users' vehicle usage into three categories and consequently presented three concept vehicles: EZ-GO for day-to-day routine activity, EZ-PRO for moving heavy products, and EZ-ULTIMO for moving heavy products with premium mobility to ride on a pleasant or special day \cite{Sharedmo62}. Future vehicle interiors are changing beyond providing smartphone-like infotainments and shifting into intelligent and innovative spaces tailored to the purpose of the vehicle. In order to provide satisfactory vehicle interior experiences to users, vehicle design should deliver a unified vehicle interior design concept for the purpose intended by its users. 

Therefore, the devices composing the vehicle interior and the sensation of space must be unified in expressing the purpose provided by the vehicle. It is essential to define the affective role of the vehicle product in delivering cohesive feelings within the vehicle \cite{norman2004emotional, norman2013design}. The dictionary definition of 'affect' is ``relating to, arising from, or influencing feelings or emotions," which refers to things naturally aroused when seeing or using an object \cite{affectve88}. Similarly, well-designed affective vehicle interiors allow users to naturally immerse themselves in the provided experience and provide an emphatic impression from such emotional engagement \cite{diamond2003love}. Therefore, it is essential to define the affective role while designing a better vehicle interior.

In order to define the affective role, related studies collect and organize adjective units by exposing users to as many designs to arouse various impressions. However, it is hard to provide various vehicle interior stimuli to expose users due to their large scales. Thus, the affective role of the vehicle interior has been identified in specific vehicle components, such as the steering wheel and clusters \cite{nagamachi1995kansei, nagamachi2004kansei}, infotainment displays \cite{yun2004affective}, or human-machine interfaces \cite{braun2021affective}. Furthermore, recent industry changes actively introduce affective solutions within the vehicle interior design. For example, Benz has presented the AVTR concept car, which tracks users' emotional state and reflects on the driving style \cite{Inspired23}. In addition, KIA introduced the real-time emotion-adaptive driving system at CES 2019 \cite{CES2019K80}. By tracking drivers' emotions, the system predicts their wants on in-vehicle experiences and provides a seat and infotainment system tailored to them. 

Nevertheless, the current affective solutions provided by vehicle interiors focus on human emotions such as anger or excitement, which are intuitive and straightforward emotions and sensations expressed by users. However, there are limitations to studies that encompass users' complicated feelings and define users' affective sensations felt within the vehicle. For example, when users are provided with a welcome light after opening the door, the instant emotion aroused from it is surprise, while the affective sensation delivered to users is a thoughtful and considerable impression. Therefore, it is necessary to identify hidden sensations of users within the scope of future vehicles and define the affective role of future vehicle interiors by organizing observed sensations.

This study aims to define the affective role of future vehicle interiors from users' expectations for the innovative future autonomous vehicles. To gather the insights, we designed focus groups among people with similar vehicle usage purposes and gathered data on expected affective roles of future vehicle's interior space through interviews. Then, we performed a content analysis of the gathered information to identify the affective role of the future vehicle interior. We expect this study to provide a framework for future research on automotive affective computing solutions in line with the current innovations in the automobile industry. Thus, the study results could provide a foundation of primary data for affective interior design and apply to an emotional evaluation scale in evaluating future design proposals. Lastly, the study is expected to support more satisfactory user experiences in developing an affective vehicle interior. 

\vspace{-3mm}

\section{Related Works}
\subsection{Affective Design for Automobile Interior}
\textit{Affective design} is a design approach that incorporates the delicate desires of users into the product design. Norman described Affection as ``Senses relating to the mind", specifically ``the action or result of affecting the mind somehow; a mental state brought about by any influence; an emotion, feeling." Therefore the product's affection could be emanated from the user's impression of the product expressed in expressive words \cite{norman2013design}. In general, affection is described at three levels. The first level is the visceral appeal, which concerns the sensibility immediately perceived visually by looking at a product. The second is the behavioral appeal, the sensibility of the pleasure and effectiveness felt when using the product. Finally, the third level is the reflective appeal, the sensibility that summarizes the product's evaluation and impression. The affective role of the product derived from these three levels is the sensibility carried out to users when using the product, and it also guides design direction during the design process \cite{khalid2004framework}.

The definition of the affective role of vehicle interiors was centered on Kansei engineering researchers; affective computing researchers have also been defining roles recently. In early studies, the role was determined by the vehicle interior components' quality and usability. Nagamachi defined the primary affective function of the center-fascia design of vehicle interiors as tight (the products perceived as high-quality products installed tightly), direct, speedy, and communicable \cite{nagamachi1995kansei}. Tomio and Kiyomi defined the main affective functions of the vehicle interior cluster area as luxurious or easy to understand \cite{jindo1997application}.

As bigger and wider display takes place in the center-fascia area of the vehicle interior, the affective role of the usability of a human-machine interface (HMI) and infotainment systems has been established in the Mercedes Benz user experience (MBUX) \cite{Mercedes58}. Willam et al. argued that HMI devices in vehicles should be safe, cognitive, non-distracting, co-present, and sociable \cite{williams2014affective}. Philip et al. have defined the affective role of the in-vehicle feedback systems as trustful and persuasive \cite{hock2016elaborating}. Finally, Lee et al. identified the affective role of the interface in the future vehicle as customizable, connective, social, maintainable, accessible, informative, spacial, user-centered, private, trustworthy, health-conscious, and secure \cite{lee2020exploring}.

In addition, the spatial aspect of the affective role of the vehicle interior has been taken into essential consideration as the purpose and usability of the vehicle diversity with the development of autonomous driving technology. Kwon and Ju distinguished the purpose of future vehicles into five categories: watching content, rest, work and self-development, personal and home management, and entertainment \cite{kwon2018interior}. Pettersson identified the emotional role of future vehicles as protective, comfortable, social, emotion regulating, liberating, innovative, relaxing, novel, and efficient \cite{pettersson2017travelling}. Though researchers have identified the affective role of the future vehicle interior, there is a lack of systematic analysis in defining and organizing each role. Therefore, a framework that can reorganize the affective role under a coordinated connotation is needed. 

\vspace{-3mm}

\subsection{Methods for Defining Affective Role}
When defining the affective role of the product design, it is common to collect responses from users by providing a sufficient number of design proposals or organizing in-depth discussion sessions regarding the design proposals. In the case of the products already in the production stage, it is advantageous to expose users to diverse concepts and situations to gather user responses and thoughts from diverse user groups. Kwon and Ju utilized factor analysis to extract the role of future vehicles after collecting user responses to the role of future vehicles. They have provided a thorough explanation with a summary of 11 kinds of published articles done by affective computing researchers and presented 18 types of concept cars created by automotive OEMs when collecting opinions from 116 users. They define the five roles of future vehicle interiors: watching content, rest, work, personal management, and entertainment \cite{kwon2018interior}. Michael et al. conducted content analysis, a qualitative analysis technique, on 131 in-vehicle affective computing solutions to determine their affective role. The roles of affective computing solutions were anger, sadness, frustration, fatigue, happiness, valence, and arousal \cite{braun2021affective}. However, there is a limitation in defining more complex sensational aspects as the affective role defined in previous studies is mainly derived from currently available functions and solutions. 

To explore the role of the future vehicle interior space, Petterson utilized a low-level prototype of an arbitrary 1:1 scale vehicle interior space created with markers and chairs. They conducted an ideation session on the prototype \cite{pettersson2017travelling}. They define the role of future vehicle interior space as safe, comfortable, caring, social, free, relaxed, new-era, novel, and efficient. 
To predict the purpose of using AR devices inside future vehicles, Weigand et al. classified them as safe, navigating, convenient, entertaining, and monitoring by conducting a focus group interview and literature research \cite{wiegand2019incarar}. 
Cha et al. recruited various vehicle users to conduct stakeholder interviews to understand their deep demands when providing a more unified user experience of both the vehicle interior and mobile devices within the vehicle. 
They determined user demands as all-in-one, seamless, limitless, productive, companion, emotionally affiliated, continuous, attractive, and responsible \cite{cha2015identifying}. 
Petterson and Karlsson suggested that the qualitative analysis done with abstract concepts of potential solutions better support exploring the expected values from participants \cite {pettersson2015setting}. 

Therefore, exploring design implications and collecting expected values of future vehicle interiors is more beneficial than providing existing solutions. 
Similarly, this research borrows the concepts from mentioned studies by conducting focus group interviews with diverse target user groups to derive the affective role of future vehicles. The expected outcome of the study is a framework format of the affective role of future vehicles when predicting new experiences and services provided within them.

\vspace{-3mm}

\section{Methods}
This study intends to define the affective role of future vehicle interior spaces by understanding user expectations. Therefore, we collected user opinions on the affective role of future vehicles by conducting a focus group interview with various user groups in anticipation of creative and constructive discussion. 

\vspace{-3mm}

\subsection{Participants}
As shown in Table \ref{tab:interviewee}, thirty-one participants were recruited through the institution's mailing list and our final sample included ten focus groups of 14 males and 17 females. As the main focus of the study is to understand user perceptions and consumer demands for future vehicles, we recruited people in their 20s who were familiar with high technology devices and services (mean of all participants' age = 28.23; standard deviation of all participants' age = 3.59). 

As user sensation changes significantly depending on vehicle ownership, we classified groups accordingly \cite{oliver1993profiles}. We recruited various college students who are interested in cars but have not yet owned a vehicle (G1), rented a car (G2), inherited (G3), and purchased and used a pre-owned car (G4). Furthermore, we included interviews with users with overseas driving experience (G5, G6). Since the user experience of vehicles is influenced by the culture, we expected to overcome the limitations of the study set up as this interview was conducted mainly with Korean participants \cite{edelmann2021cross}. Moreover, we recruited participants closely related to the automotive industry: graduate students working on car-related projects (G7) and project managers of the car industry (G8). Furthermore, we also invited experts from other industries (G9) to hear different opinions on the current trend of future vehicles. Lastly, we recruited users who mostly use a vehicle with their child (G10) to observe the purpose of using the vehicle while caring for other people.

\begin{table}[h]
\vspace{-2mm}
\caption{Demographic data of focus groups (N=10) and its participants (N=31)}
\vspace{-2mm}
\label{tab:interviewee}
\resizebox{\textwidth}{!}{%
\begin{tabular}{lll}
\toprule
\textbf{Group} & \textbf{Description} & \textbf{Participants (gender: M for male, F for female, age)} \\ \midrule
G1    & 20s college students have an interest in cars & P1.1 (M, 24), P1.2 (M, 23) \\ \midrule
G2    & 20s office workers frequently using car-sharing service & P2.1 (F, 26), P2.2 (F, 25), P2.3 (M, 24), P2.4 (F, 29) \\ \midrule
G3    & 20s students inherited their parents' car & P3.1 (F, 27), P3.2 (F, 25), P3.3 (F, 29) \\ \midrule
G4    & 20s students bought an used car & P4.1 (M, 29), P4.2 (F, 28) \\ \midrule
G5    & 20s students have driving experience abroad (Kazakhstan and China) & P5.1 (F, 28), P5.2 (F, 28)\\ \midrule
G6    & 20s students have driving experience in USA & P6.1 (F, 29), P6.2 (M, 29), P6.3 (M, 31) \\ \midrule
G7    & 20s graduate students had project experience with car company & P7.1 (M, 32), P7.2 (F, 29), P7.3 (F, 29) \\ \midrule
G8    & 20s office workers working in management team for car company & P8.1 (F, 29), P8.2 (M, 25), P8.3 (M, 29), P8.4 (M, 27) \\ \midrule
G9    & 20s office workers working in management team (not car company) & P9.1 (M, 24), P9.2 (M, 27), P9.3 (M, 28), P9.4 (M, 30) \\ \midrule
G10   & 30s working moms usually driving with their young toddler & P10.1 (F, 34), P10.2 (F, 38), P10.3 (F, 36), P10.4 (F, 35) \\ 
\bottomrule
\end{tabular}%
}
\vspace{-5mm}
\end{table}

\subsection{Procedure}
\begin{figure}[h]
    \centering
    \includegraphics[width = .5\textwidth]{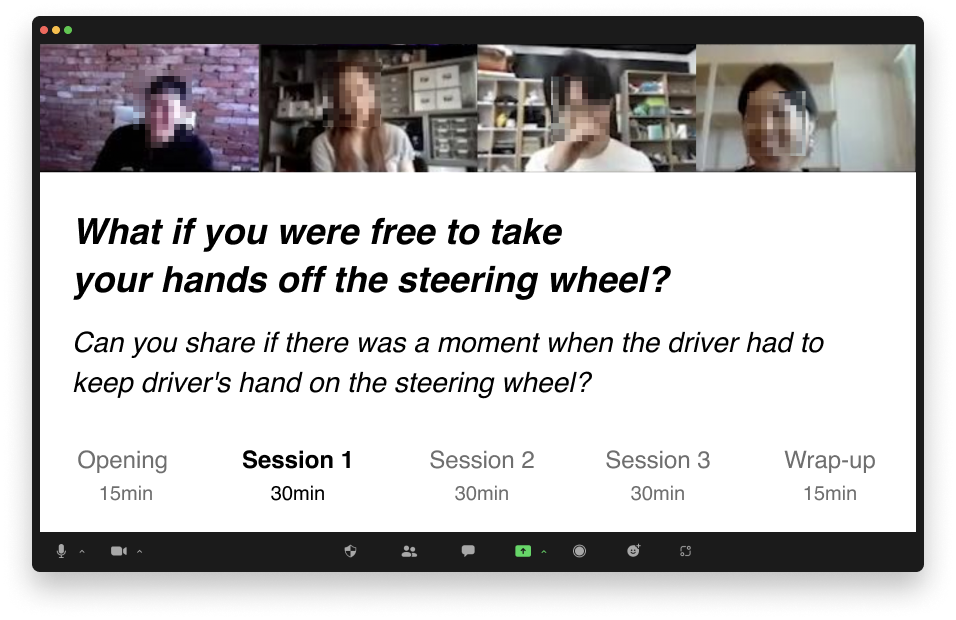}
    \vspace{-5mm}
    \caption{Focus group interview using zoom (Example of Group 7, Session 1)}
    \label{fig:zoom}
    \vspace{-7mm}
\end{figure}

The interview was conducted via the online conference program Zoom due to the COVID-19 outbreak, as shown in Figure \ref{fig:zoom}. The interview was divided into five stages: an opening and ice-breaking session, three main sessions, and a warp-up session. The entire session lasted about 2 hours. Each interview was conducted as a group interview session. In the session, users could quickly suggest ideas, empathize with others by explaining experiences of encountering difficulties in current vehicle usage, and suggest expected solutions to the difficulties they felt in future vehicles. Then, further discussions on proposed solutions and prospective experiences from future vehicles took place. The entire interview session was conducted as an open discussion without any interruption by the interviewer. Moderators took notes for the interview and summarized the event for wrap-up.

The opening session was the ice-breaking session that induced active communication among the interviewees by sharing participants' understanding of the autonomous vehicle and their perceptions. 
The main sessions were structured into three major sessions about significant changes expected with autonomous vehicles:
1) Allows the driver to take their hands off the steering wheel,
2) Blurs the boundary between the driver and passengers and enables communication within the vehicle,
3) The design freedom for the interior space of the vehicle is significantly increased.  
Main sessions were composed of two main questions; 15 minutes was soent on each question. According to each question, we requested that participants use their imagination (Imagine the moments and summarize on the chat room) for 5 minutes, and share (Sharing the moments and providing similar situations to others) for 10 minutes to encourage participants to describe the future vehicle's expected sentiments and demands during the interview. Finally, we wrapped up with the final interview. The reward provided to each participant was 30\$. The detailed structure of the interview questions is as follows.

\begin{itemize}
    \item Ice breaking session (15min)
        \begin{itemize}
            \item What kind of vehicle do you own (or can be purchased if you do not own it), and can you share some regrettable moments while driving the vehicle?
            \item If you could attach new options to your current vehicle, which option would you want to add? 
            \item Or if you can buy a new vehicle which, kind of car you want to purchase?
        \end{itemize}
    \item Session 1. What if you were free to take your hands off the steering wheel? (30min)
        \begin{itemize}
            \item Can you share if there was a moment when the driver had to keep their hand on the steering wheel?
            \item If the advent of autonomous driving technology frees the hands of the driver, what kind of experience do you expect the vehicle to provide?
        \end{itemize}
    \item Session 2. What if free communication becomes possible due to the blurring of the boundary between driver and passenger? (30min)
        \begin{itemize}
            \item Can you share a moment that was disappointing in communication during the in-vehicle experience because the driver's concentration on driving had to be considered first?
            \item What kind of vehicle interior do you expect to be provided if it can be changed so that there is no problem with communicating inside the vehicle with the advent of autonomous driving technology?
        \end{itemize}
    \item Session 3. What if the design restrictions on the interior of the vehicle were completely removed so that you could freely mount the options you want? (30min)
        \begin{itemize}
            \item Can you share a moment when it was inconvenient to have certain components attached to the vehicle in a certain space?
            \item Given that the interior components of a vehicle can be freely moved or changed, what kind of vehicle space do you think will be provided?
        \end{itemize}
    \item Wrap-up session (15min)
     \begin{itemize}
            \item (With showing the chat room with summarizing) The affective experiences that came out in today's interview are as follows, so could you tell me if there's a topic that you'd like to discuss more?
            \item Lastly, if you don't have enough time or if there's a topic you want to handle more don't feel hesitate to share and after sharing final feedback, we'll end the session.
        \end{itemize}
\end{itemize}

\subsection{Data Analysis}
We performed content analysis with four rounds of the coding process for data analysis. For the preprocessing of the data, all interview data were transcribed and transferred into 437 quotes. Two researchers conducted open coding with 409 quotes for the first round of content analysis. Then, they collected similar quotes and clustered them into 232 factual levels. Finally, two researchers conducted closed coding to divide clustered factual levels into two categories: facts related to overcoming current pain points and expectations for a new experience with future vehicles. Two researchers divided the collected facts into two categories for open coding: overcoming pain and expecting new experiences. Clustering of the collected facts was to support researchers in extracting emotional vocabulary. The open coding session resulted in 130 emotional facts on overcoming pain and 102 emotional facts on expectations for new experiences. Then, the closed coding was conducted by each researcher on clustered emotional facts. Finally, the initial code table was designed after rounds of discussion of mismatching codes from the results of the closed coding session. We invited two other researchers for an intercoder evaluation of the initial code table. We finalized the code table with the final themes, as shown in Table \ref{tab:codes}.

\begin{table}[h]
\caption{Final themes and affection codes}
\label{tab:codes}
\vspace{-3mm}
{%
\begin{tabular}{lll}
\toprule
\textbf{Theme}       & \textbf{Code}         & \textbf{Description} \\ \midrule
Exclusiveness & Focused (23)      & \begin{tabular}[c]{@{}l@{}}Users would like a convenient and thoughtful place for routine \\ activities or focus \end{tabular}  \\ \cmidrule{2-3}
              & Stimulating (21)    & \begin{tabular}[c]{@{}l@{}} Users need for more personalized and explicit interior space \end{tabular}\\ \cmidrule{2-3}
              & Amused (24)       & \begin{tabular}[c]{@{}l@{}} Users recognize as a unique space that makes enjoyable and amenable to\\  share pleasant activities with others \end{tabular} \\ \cmidrule{2-3}
              & Pleasant (21)       & \begin{tabular}[c]{@{}l@{}} Users expect experience to be akin to chauffeured vehicle  \end{tabular} \\ \midrule
Convenience   & Safe (11)         & \begin{tabular}[c]{@{}l@{}} Users expect safe and protected sensation  inside future vehicle. \end{tabular} \\ \cmidrule{2-3}
              & Comfortable (46)  & \begin{tabular}[c]{@{}l@{}} Users expect future vehicle interiors to deliver caring emotional,\\ personalized services, and reliable solutions  \end{tabular}   \\ \cmidrule{2-3}
              & Accommodated (33) & \begin{tabular}[c]{@{}l@{}} Users expect to have an accurate and immediate understanding of \\ vehicle status and information \end{tabular}   \\ \cmidrule{2-3}
              & Organized (52)    & \begin{tabular}[c]{@{}l@{}} Users expect to future vehicles to provide efficient and customizable space \end{tabular}  \\ \bottomrule
\end{tabular}%
}
\vspace{-7mm}
\end{table}

\section{Results}

The users expect the interiors of autonomous vehicles to primarily fulfil two affective roles: advanced exclusiveness and advanced convenience. The identified affective roles of each aspect are a total of eight visceral levels, four visceral levels each. Under the advanced exclusiveness aspect, prominent affective roles are focused, stimulating, amused, and pleasant. Under the advanced convenience, safe, comfortable, accommodated, and organized.

\vspace{-2mm}

\subsection{Role of Advanced Exclusiveness of Future Vehicle}

\subsubsection{Focused}
Users expect to have focused sensibility from future vehicle. Participants reflected that the current role of experiencing focused sensation was solely on driving the vehicle. For this reason, they expected to have a more safe and comfortable time enjoying the daily activity on the road within the autonomous vehicle in a more considered environment. For example, G2 and G9 anticipated having an office-like space within the vehicle where they could check e-mails and do simple work while commuting. In addition, almost all female participants expected to have a convenient and comfortable place to put on makeup. We observed that users would like a convenient and thoughtful place for routine activities or focus. 

Participants explained that the vehicle should understand user activities and provide and convey relevant sensations within the future vehicles. For example, G2 expressed that \textit{"I think it is most important to know what I am trying to do and gently guide me to do that activity naturally. I do not do it because I am too lazy to do it. After all, it is annoying"} about the office-like environment. In addition, participants stated that they expect to feel treated and served by the vehicle as they experience the interior that provides a tailored and focused atmosphere by understanding the circumstances of the environment.  

\vspace{-3mm}

\subsubsection{Stimulating}

Users are hoping to have a more stimulating and personalized interior environment. The current user experience of achieving stimulating sensation from the vehicle interior is to purchase additional options on the vehicle by modifying premium car seats or upgrading to Hi-Fi audio. However, user responses to such additional options were similar to what is provided by the automotive OEMs (G1) or after-market purchases as it is too expensive (G3, G4). They expect to be accommodated with more freedom to personalize interior components and decorative devices within the future vehicle interior. G1 and G3 proposed vehicles that could replace vehicle devices with expensive options in the future, while G8 proposed replaceable vehicle seats.

G8 expressed in the interview that \textit{"I may be able to stylize the interior to my personal taste. For example, if I replace seats with different designs, interior devices, and displays that change their moods and colors accordingly."} It explains a need for accommodation within the future vehicle with more automated stimulating options to provide personalized and explicit interior space.  

\vspace{-3mm}

\subsubsection{Amused}
Users expect the amusing role of vehicle interiors with greater customizability for its users. Participants mentioned that the entertainment experience of the current vehicle interior is limited to watching content or listening to music. On the other hand, G6 expected future vehicles to prove an immersive \textit{``theater-on-wheels"} for long road trips. G10 hoped to have a shared video game studio with friends after giving children a ride to school.

Especially when G10 explained about the video game studio, they stated that \textit{``After sending the children to school, there is a bit of time left, and in the meantime, it would be nice to invite other parents to play games together. So it is to provide a mingle place with fun activities or contents to enjoy."} Furthermore, it explains that the vehicle will be remembered as a unique space that allows users to share pleasant and enjoyable activities with others.

\vspace{-3mm}

\subsubsection{Pleasant}
Users expect a more pleasant emotional experience with advanced technology from future vehicles. With autonomous driving technology, users can enjoy the scenery on the road; thus, interactivity with the the outside and the sensation of openness within the vehicle would be more critical. For example, G3 expected to see the night sky on a gloomy day. G7 hoped to see the sky on a rainy day while driving. Similarly, G5 and G6 suggested the interior space where they could observe aurora or star lights from the autonomous vehicle. Finally, G11 suggested a system to communicate easily with other vehicles on the road. 

Many participants collectively illustrated glass-like interiors or open spaces that assisted the openness sensation achieved from enjoying the scenery, which is a luxurious experience as it is similar to a chauffeured vehicle. They also described that enjoying swift moving on the autonomous vehicle would provide different sensations as users will enjoy different activities inside the vehicle rather than focusing on driving. 

\subsection{Role of Advanced Convenience of Future Vehicle}

\subsubsection{Safe}
Users expect to have future vehicles communicate the feeling of safety within the autonomous vehicle. During the interview, many participants expressed that drivers and passengers would feel uncomfortable as they have less control and focus on driving with autonomous vehicles. As they feel more alerted and cautious with the actual driving situation from the current vehicle experiences, they would like to feel similarly secure within the future vehicle.
G2 responded to feeling secure inside the future vehicle as \textit{``If you use an autonomous vehicle, you will hardly care about the road situation, but I think it will be essential to feel safe and secure in the future vehicle. Moreover, I would be more surprised if I am not prepared for accidental moments like sudden breaks or interruptions from other cars."} It implies that continuously delivering a safe and protected sensation inside the future vehicle is critical.

G2 also discussed possible solutions for conveying a sense of safety in future vehicles. They explained, \textit{``So I think it will be very important for the vehicle to quickly and accurately judge the moment in advance and guide it immediately. Then I think I'll be able to trust the car when I ride it, so I'll always want to ride this car."} Their comments imply that accurate and advent notification of any accidental circumstances is critical. Delivering a continuous sensation of feeling safe and secure with responsive notification of any accidental circumstances would provide more trust in the vehicle system.

\vspace{-2mm}

\subsubsection{Comfortable}
Users expect to have comfortable and convenient space in future vehicles. Current vehicles are designed with ergonomics and safety regulations tuned to driven cars. For future vehicles, users expect to have space for resting comfortably (G10) and sleeping while the vehicle is running on the road (G2). In addition, users highlighted that future vehicle interior space should provide necessary convenience and a pleasant place.  

Additionally, G2 added that comfortable mattress-like seats might be helpful while commuting. G10 stated that convenient devices should aid babysitting. We found out from the interview that a future vehicle interior should deliver caring emotion, personalized services, and reliable solutions to users.

\vspace{-2mm}

\subsubsection{Accommodated}
It is essential to have accommodated spatial experiences that are more personalized and tailored to users. During the interview, G6 described that \textit{``It feels like a future autonomous vehicle would be a personal salon with a private secretary as it provides necessary assistance tailored to my schedule, emotions, and driving style. Moreover, it would be more reliable and thoughtful if the vehicle provided personalized services from when I hop into the vehicle until I get out of it."} It points out that a caring and personalized vehicle experience is essential for future vehicles.

Based on the interview results, users expect to have an accurate and immediate understanding of their status and information from future vehicle interiors. Furthermore, such an understanding of users implies that users hope to be served by the vehicle with more reliable and tailored services.

\vspace{-2mm}

\subsubsection{Organized}
Users expect to have a space that provides convenience and easily manipulable aspects.
Users feel that the vehicle's space is limited compared with other personal living spaces or transportation systems. Current vehicles are packed with necessary components that their users manipulate, and such components are an inevitable necessity in the interior space. Users expect to have more space with the future vehicle as electrification and the autonomous driving system eliminates the required components previously required. This change will provide more organizable space for users and supports various purposes. In addition, users expect more compatibility of the vehicle interior space for their usage. 

From the interview, G3 responded, \textit{``I bought an electric car because it was roomier and taller than other vehicles. Frequently, I am packed with many things in my car. Then, if the car is not tall enough, it looks stuffed and squashed. I cannot even check cars behind me. For this reason, I would consider more room provided inside the vehicle and the efficiency of the interior space."}
Providing efficient and manipulable space for users from future vehicles is essential even though the ultimate space is limited. In addition, conveying the sensation that users are in control of the space in its organization would be considered.

\vspace{-2mm}

\section{Discussion}
In this study, we defined the affective roles of autonomous vehicle interiors by collecting users' emotional expectations from focus group interviews. This research has defined future autonomous vehicle interior space with four affective roles of exclusiveness and four affective roles of convenience. All together, the eight roles are focused, stimulating, amused, pleasant, safe, comfortable, accommodated, and organized.

Two essential aspects, exclusiveness and convenience, were derived, similar to previous research that determined perceptions of devices in the future vehicle interior. William et al. revealed that devices presented within future vehicle interiors deliver the contextual acceptability of future vehicle interiors and resolve impaired driving for a safe driving experience \cite{wellings2010understanding}. The overall emotional response and user perception conveyed by devices is similar to the exclusiveness role identified by our study that provide luxurious emotional perception. Accordingly, providing safe and comfortable autonomous vehicle driving is similar to our research's convenience. More specifically, our study's focused, stimulating, amused, safe, comfortable, and accommodated roles were similar to previous studies. Previous studies have identified that factors such as intention, influence, fun, trust, facilitation, and expectancy are essential in describing how people perceive autonomous vehicles and their acceptance \cite{osswald2012predicting, rodel2014towards, hewitt2019assessing}. These resemblances in factors imply that people's perception of the autonomous vehicle will likely affect their emotional response to the autonomous vehicle interior. 

Furthermore, previous studies identified roles that influence users' emotional experience of the autonomous vehicle interior. Cha et al. proposed six types of emotional perceptions of integration of mobile devices into vehicles \cite{cha2015identifying}. Our study revealed two additional factors, such as amused and accommodated, from the six of Cha's study when defining the affective roles of autonomous vehicle interiors. Therefore, the eight roles described in our study are to fulfill the emotional aspects conveyed from the interior context, which involves overall aspects of spatial atmosphere within vehicle interiors rather than mobile device integration.  

Petterson has studied users' emotional experiences in the autonomous vehicle interior \cite{pettersson2017travelling}. His study has indicated these emotional aspects into nine roles: care, new-era, novel, social, free, relaxed, safe, comfortable, and efficient. Petterson conducted the study by implementing abstract prototyping of the autonomous vehicle interior space. As a result, nine roles identified by Petterson are to deliver the expected emotional aspects of its prospective users. Though most of the emotional roles are similar to the eight roles identified in our study, the results of our study imply that accommodated aspects of autonomous driving should be considered when defining the emotional experiences caused by the future vehicle interior. 

\vspace{-5mm}

\subsection{Implications on Design Ideation}
The identified affective roles could provide a foundation for developing future vehicle interior design and the conceptual direction of the vehicle. For example, as the target user groups and expected concepts of the vehicles are identified, the affective role of the vehicle concept could be explained with the eight roles identified in this study. Moreover, the vehicle concept's selected roles and emotional aspects could be further expanded into a design brief and implemented into an overall impression of the interior and components creating related sensations. Especially when it is presented with vehicle interior context, this could remind the role of providing relative affective role with components such as seats, color material and finishing, human-machine interaction devices, adaptive music, ambient light, empathic speech, and infotainments \cite{braun2021affective}.

\begin{figure}[b!]
 \vspace{-7mm}
    \centering
    \includegraphics{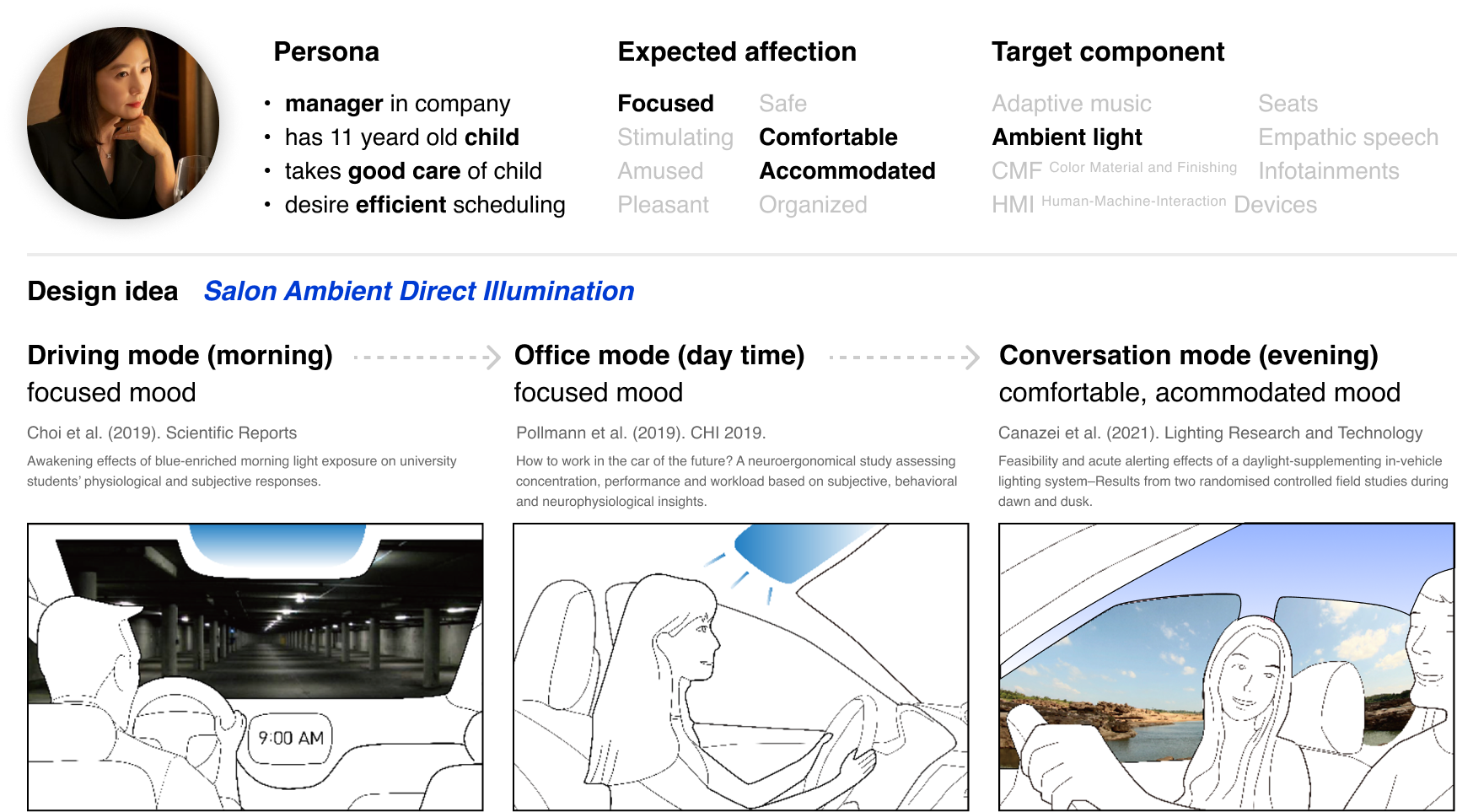}
    \caption{Exploring vehicle interior lighting based on affective roles defined related to persona's expected experience.}
    \label{fig:lightdesign}
\end{figure}

The implication of the research results was achieved by applying them in deriving a design concept for an affective ambient lighting solution for the autonomous vehicle interior, as shown in Figure \ref{fig:lightdesign}. We selected our intended user as a career mom who is outgoing and passionate about her work while taking good care of her child. A detailed explanation of our target user is as follows:
\vspace{-1mm}
\begin{itemize}
    \item She is a manager of the company with ten years of experience.
    \item She has an 11-year-old child who attends elementary school.
    \item She takes good care of her child. 
    \item She wants efficiency in her scheduling as she has limited time for herself.
    \item She expects to have focused, comfortable, accommodated roles in the vehicle.
\end{itemize}
 \vspace{-1mm}

Based on this consideration, the vehicle salon ambient direct illumination design is as follows. 
The designer draws a scenario in which the user uses the designed vehicle lighting. Affective roles were matched on the completed scenario to provide satisfying solutions to users, and evidence-based designs using related affection studies were applied. First, when the driver enters the car, blue-enriched awakening lights are provided for driving on the way to work. It creates an environment where drivers can focus on driving while waking up from sleep \cite{choi2019awakening}. During the day time, an emotional lighting solution helps her focus on work within the vehicle by providing brighter and greener lighting to make it easier to do tasks \cite{pollmann2019work}. Next, when a child gets into a car after school, it provides appropriate bright white lighting to make it easier to talk about the day, helping them communicate in the vehicle \cite{canazei2021feasibility}. The presented lighting solution provides a comfortable sensation with familiar materials of the lighting elements of a living room. The lighting solution also provides a focused emotional experience when communicating or working inside the vehicle. Furthermore, an accommodating sensation is provided with the natural transition of the lighting configuration depending on user activity. 

The proposed framework of the study and identified affective roles can provide a fundamental guideline in design ideation and inspiration to explore new ideas in providing a solution for the vehicle interior for satisfying users' emotional experiences. According to the emotional design process summarized by Triberti et al., it is important to clearly define the affective role of users' expectations and start to explore the design based on the defined sensibility \cite{triberti2017developing}. After that, when the defined sensibility and the design target are clarified, it provides a research lens to search users' responses between design components and affection in affective computing research. Then, as Desmet et al. describe in their book, an evidence-based design process could provide the best emotional design to users \cite{desmet2007emotional}.

\vspace{-2mm}
\subsection{Implications on Design Evaluation}

The affective role of the vehicle derived in this study can be used as data to evaluate the emotional value of existing products and discuss the direction of improvement to the future vehicle interior with the specific emotional experience and enhancement.
When evaluating the vehicle's design, the current vehicle interior spaces can be mapped by clearly understanding the emotional aspects of the eight derived affective roles of the study. Based on the mapped results, current deficiencies or areas that need improvement can be reviewed, and the future direction of design development can be derived relatively.
In addition, the amusement affective role identified in the study could be added strategically to express design direction more delicately.

\begin{figure}[b!]
\vspace{-3mm}
    \centering
    \includegraphics{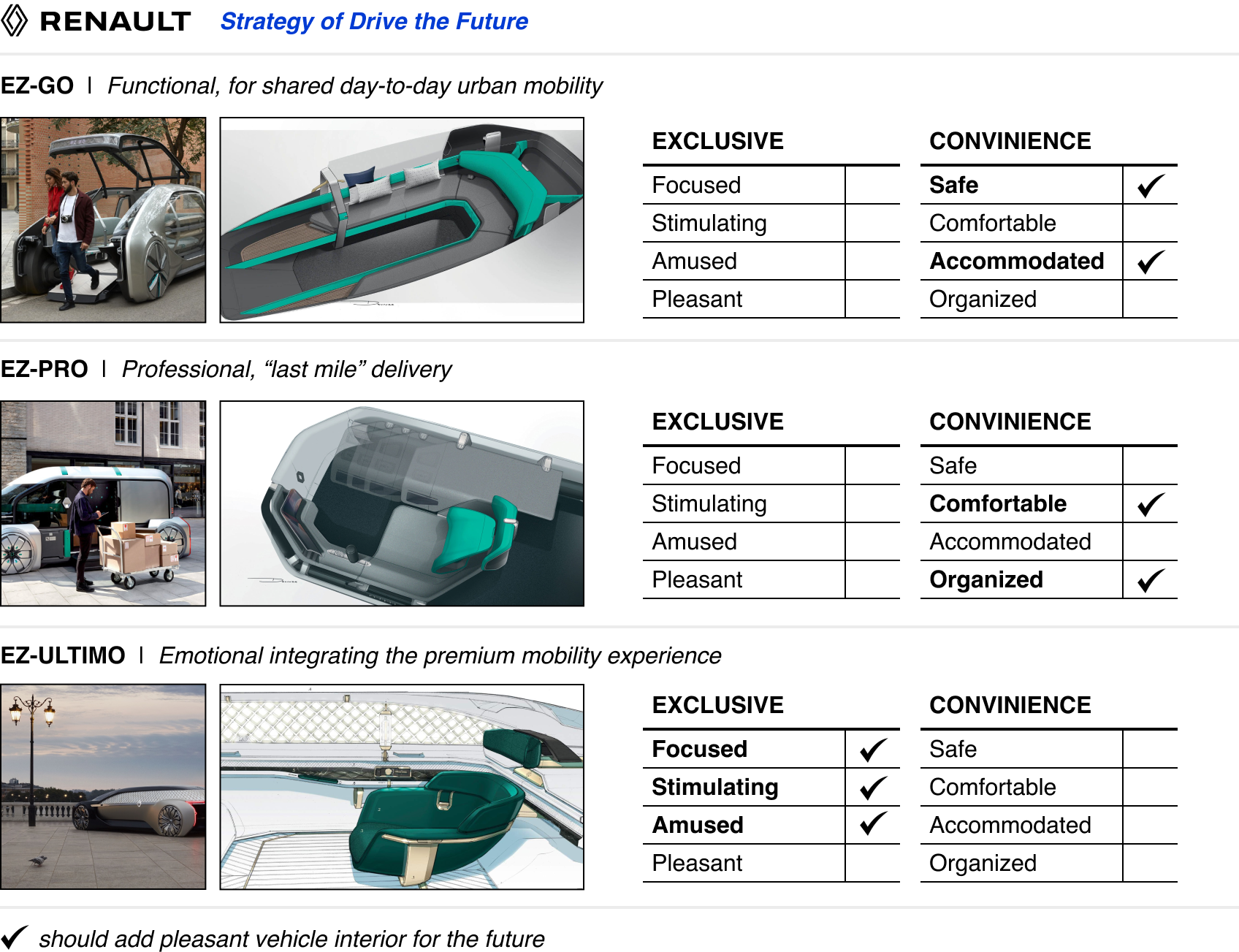}
    \vspace{-2mm}
    \caption{Evaluating Renault's Trilogy concepts \cite{Sharedmo62} based on affective roles defined in the research.}
    \label{fig:reuanlt}
\end{figure}

As an example of utilizing the study results on design evaluation, as shown in Figure \ref{fig:reuanlt}, we projected Renault's Trilogy concepts onto identified affective roles to draw points of design improvements from the strategic design point of view \cite{Sharedmo62}.
EZ-GO, the first concept of the Trilogy concept, is a concept that focuses on efficiently allowing users to move around the city center. Thus, EZ-GO can be evaluated based on the affective role by conveying the accommodated sensation that allows people to move around places efficiently and emphasize the safe feeling from any road conditions in the city center.
The second concept, EZ-PRO, focuses on delivering goods from place to place. Accordingly, it is vital to convey an organized sensation when loaded with goods and a comfortable feeling to make the interior space convenient to use.
The last concept, the EZ-ULTIMO, is a super-personalized, luxury self-driving car. EZ-ULTIMO has a stimulating sensation in a luxury interior component and a focused feeling to support people's activities inside the vehicle. 
Based on the mapping results, we have confirmed that Renault's Trilogy concepts and its design direction could easily be explained with the affective roles identified in the study.

Therefore, the results of this study are expected to be used as a rationalized basic knowledge to systematically organize existing vehicle design concepts and seek improvement directions on the user's emotional experience. Simply, user responses can be collected on a Likert scale using defined affective roles from this research and used as an indicator for quantitatively measuring users' emotional experiences, for example, with a user experience questionnaire \cite{laugwitz2008construction}, NASA-TLX \cite{hart1988development}, and PANAS \cite{watson1988development} which are widely used to quantify users' emotional experiences from using products. In this circumstance, it is expected to make a clear contribution to automotive practice and academia.

\subsection{Limitation}
Our study collected data with limited diversity; we gathered verbal data from interviews using an online conference program. Though the verbal data included fruitful user expectations on affective vehicles, further research can complement the affective role concepts by collecting additional sketch data that clarify user intentions and help organize design elements of vehicles for affective design. Also, this study performed analytic coding processes that matches behavioral and reflective codes with visceral codes. Thus, there is a problem that behavior and reflective codes are difficult to have meaning independently without linkage with visceral codes. If the in-depth study and analysis of behavior and reflective codes can be performed in future work, it would be clearer to set the direction of the idea based on each affective role, and the content of specific design evaluation scales related to the role could also be derived. Our study tried to cover various age groups as much as possible, but the experiment mainly consisted of consumers in their 20s. The impression and definition of future vehicles may differ for different age groups, affecting user expectations of the affective role provided by the vehicle interior. Therefore, in future studies, it is expected that it will be possible to define a more systematic influential role by conducting experiments on various generations or grasping the technical socio-cultural characteristics expected from future vehicles divided by generation.

\section{Conclusion}
This study presented the affective role of future vehicle interiors equipped with autonomous driving technology and advanced infotainment contents. The analyzed results of focus group interviews with 31 participants in 10 groups indicated that the novel affective roles can be defined based on the advanced exclusiveness and convenience of future vehicles. Eight affective roles described in visceral levels included focused, stimulating, amused, pleasant, safe, comfortable, accommodated, and organized. Our results add to the understanding on the design of future vehicle interiors by clarifying the aspects that users would value to achieve a richer affective experiences. We believe that this study can contribute to the discussion on the affective design of autonomous vehicles, particularly when developing conceptual directions geared towards the affective aspects and evaluating its impact on user experiences.

\bibliographystyle{ACM-Reference-Format}
\bibliography{sample-base}

\end{document}